# Multi-Wavelength Machine Learning for High-Precision Colorimetric Sensing

**Majid Aalizadeh [1,2], Chinmay Raut [3], Ali Tabartehfarahani [1,2,4,5] and Xudong Fan [1,2,4,5,*]**

[1] Department of Biomedical Engineering, University of Michigan, Ann Arbor, MI 48109, USA; maalizad@umich.edu (M.A.); afarahan@umich.edu (A.T.); xsfan@umich.edu (X.F.)
[2] Biointerfaces Institute, University of Michigan, Ann Arbor, MI 48109, USA;
[3] Department of Computational Medicine and Bioinformatics, University of Michigan, Ann Arbor, MI 48109, USA; craut@umich.edu (C.R.)
[4] Center for Wireless Integrated MicroSensing and Systems (WIMS2), University of Michigan, Ann Arbor, MI 48109, USA;
[5] Max Harry Weil Institute for Critical Care Research and Innovation, University of Michigan, Ann Arbor, MI 48109, USA;

* Correspondence: xsfan@umich.edu;

### Highlights

Conventional colorimetric sensing methods typically rely on signal intensity at a single wavelength, often selected heuristically based on peak visual modulation. This approach overlooks the structured information embedded in full-spectrum transmission profiles, particularly in intensity-based systems where linear models may be highly effective. In this study, we experimentally demonstrate that applying a forward feature selection strategy to normalized transmission spectra, combined with linear regression and ten-fold cross-validation, yields significant improvements in predictive accuracy. Using food dye dilutions as a model system, the mean squared error was reduced from over 22,000 with a single wavelength to 3.87 using twelve selected features, corresponding to a more than 5,700-fold enhancement. These results validate that full-spectrum modeling enables precise concentration prediction without requiring changes to the sensing hardware. The approach is broadly applicable to colorimetric assays used in medical diagnostics, environmental monitoring, and industrial analysis, offering a scalable pathway to improve sensitivity and reliability in existing platforms.

### Abstract

Conventional colorimetric sensing methods typically rely on signal intensity at a single wavelength, often selected heuristically based on peak visual modulation. This approach overlooks the structured information embedded in full-spectrum transmission profiles, particularly in intensity-based systems where linear models may be highly effective. In this study, we experimentally demonstrate that applying a forward feature selection strategy to normalized transmission spectra, combined with linear regression and ten-fold cross-validation, yields significant improvements in predictive accuracy. Using food dye dilutions as a model system, the mean squared error was reduced from over 22,000 with a single wavelength to 3.87 using twelve selected features, corresponding to a more than 5,700-fold enhancement. These results validate that full-spectrum modeling enables precise concentration prediction without requiring changes to the sensing hardware. The approach is broadly applicable to colorimetric assays used in medical diagnostics, environmental monitoring, and industrial analysis, offering a scalable pathway to improve sensitivity and reliability in existing platforms.

## 1. Introduction

Colorimetric sensing is one of the most widely used methods for chemical and biological detection because it translates molecular interactions into visible changes in light absorption or transmission [1-8]. Its simplicity, low cost, and compatibility with both laboratory and point-of-care [9] settings have made it essential in applications ranging from clinical diagnostics [10, 11] and environmental monitoring [12, 13] to food safety [14-17] and industrial process control [5, 18]. Traditional approaches, however, typically rely on modeling intensity at a single wavelength from a full spectrum of wavelengths [19]. These wavelengths are often selected heuristically based on visual inspection of the spectrum, such as choosing the point of maximum absorbance. In practice, such single-wavelength readouts are often analyzed with one-dimensional fitting models such as linear regression or nonlinear curve fits like the four-parameter logistic (4PL) equation [20-24]. While convenient, this one-dimensional strategy discards most of the structured information contained in full spectra and therefore limits sensitivity, precision, and robustness. As a result, despite decades of development, many colorimetric platforms remain constrained by the same data-reduction step that oversimplifies what are inherently high-dimensional signals.

Our earlier studies addressed this limitation by showing that spectra should be treated as structured, high-dimensional datasets, regardless of the specific sensing platform [25, 26]. In the first study, we focused on resonance-based biosensing, where conventional approaches track a single resonance peak shift to estimate analyte concentration or refractive index changes [27-29]. Some resonance-based platforms also support both TE and TM polarizations, yielding distinct peaks whose joint tracking enhances sensitivity and reduces noise through internal referencing [28, 30, 31]. We demonstrated that this single-peak method overlooks the fact that multi-resonance structures contain several distinct resonances, each carrying partially independent information. By applying ridge regression to a set of resonances traced simultaneously in the absorption spectrum of a periodic silicon triangular nanorod meta-array, we showed that combining multiple resonance shifts in a ridge regression model provides far higher accuracy than tracing any single resonance alone, up to three orders of magnitude. In other words, we showed that using multiple predictors instead of just one, results in much higher accuracy in the prediction of the target variable. It was also systematically shown that by gradually adding to the number of predictors, the accuracy is consistently enhanced [25]. Although this meta-array was used as an example system, the principle is general: any resonant spectrum with multiple features can benefit from multi-feature modeling rather than one-dimensional fittings.

In the second study, we extended the approach to full-spectrum modeling [26], comparing resonance-based spectra with sharp nonlinear peaks versus intensity-modulated spectra with smooth variations. Using the same nanorod geometry as the previous work, Si produced multi-resonance spectra while Ti produced smooth intensity-based spectra, enabling direct comparison with single-wavelength methods. We found that intensity-modulated spectra exhibit a near-linear relationship with concentration at a given wavelength, allowing linear regression to reduce the mean squared error (MSE) by over 8,000-fold, whereas resonance-based spectra showed limited improvement due to nonlinear peak shifts. These results show that modeling effectiveness depends on spectral characteristics and that significant precision gains can be achieved through multi-wavelength modeling without hardware changes.

Building on this foundation, the present study provides an experimental validation in colorimetry, a domain defined by intensity-based spectra that is highly compatible with linear modeling. Using food dye dilutions as a model system for measurements, we apply forward feature selection combined with linear regression and cross-validation. We show that just twelve wavelengths are sufficient to preserve the essential concentration-dependent structure of the full spectrum, reducing the MSE by more than 5,700-fold compared to single-wavelength analysis (when using the best selected single wavelength). This result demonstrates that interpretable machine learning can substantially improve the accuracy of colorimetric assays without changes to hardware, and more broadly, that the same principles extend beyond optics to any sensing modality where structured spectral information is available.

## 2. Materials and Methods

### 2.1. Experimental Setup

Figure 1 presents a detailed overview of the experimental setup used for capturing full-spectrum transmission data from liquid-phase colorimetric samples. As shown in Fig. 1(a), the schematic illustrates the optical path beginning with a broadband light source, which emits a continuous spectrum across the visible range. The light first passes through an optical aperture, which spatially filters the beam to define its profile and suppress stray light. It then travels through a series of lenses that shape and collimate the beam, ensuring uniform propagation through the optical axis and consistent illumination of the sample region.

The colorimetric sample is contained within a standard microcentrifuge tube filled with dyed liquid and mounted vertically in the beam path. This configuration enables direct transmission measurements through the liquid column. As the beam traverses the sample, different spectral components are absorbed to varying degrees depending on the dye concentration, resulting in a wavelength-dependent attenuation. The remaining transmitted light is collected by a fiber-coupled spectrometer positioned on-axis downstream of the sample. The spectrometer records the full transmission spectrum from each sample, which serves as the raw input for all subsequent analysis and machine learning modeling.

Figure 1(b) shows a photograph of the physical system assembled on an optical breadboard. The components are mounted using standard post holders and translation stages, providing mechanical rigidity and fine control over alignment. The modular layout facilitates clear optical access to each section of the setup, particularly the sample region, while maintaining stable geometry across repeated measurements. Adjustable lens mounts allow the beam to be precisely collimated and focused for optimal transmission through the vial. An LS-1 tungsten halogen lamp is used as a broadband light source. An Ocean Optics USB2000 spectrometer is used for recording the spectra using the OceanView 2.0 software.

A magnified view of the sample holder is shown in Fig. 1(b), highlighting the central alignment of the microcentrifuge tube within a 3D holder. This holder secures the vial in a fixed position along the beam axis, ensuring that the incident light is transmitted directly through the colored liquid without significant scattering or misalignment. The surrounding hardware is designed to enable rapid sample exchange while preserving alignment consistency across the dataset. Each measurement is repeated three times by removing and placing the sample vial back in the holder, and the result is averaged out in an attempt to remove the manual measurement errors.

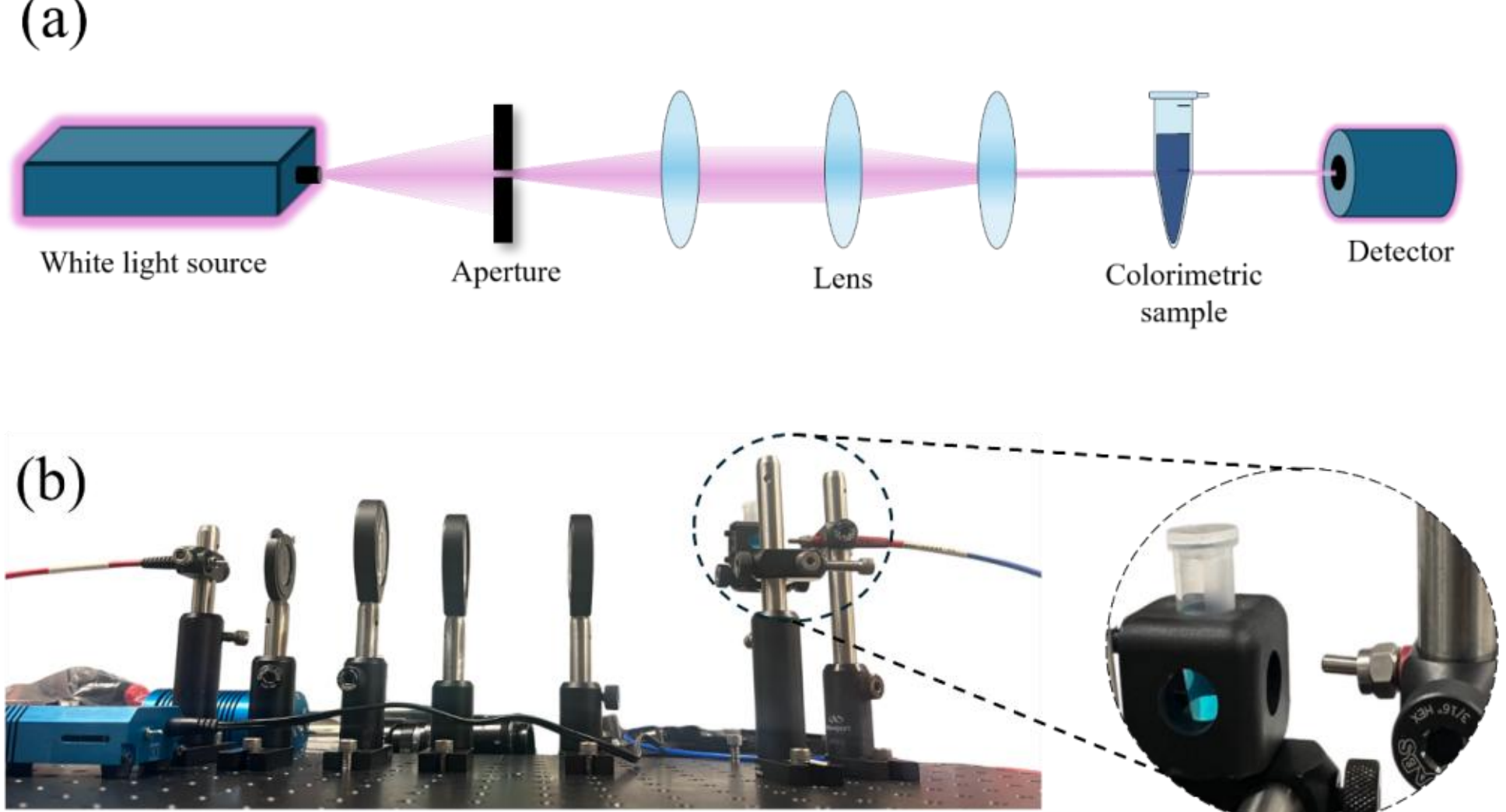

**Figure 1:** Experimental setup for transmission-based colorimetric measurements. (a) Schematic diagram of the optical path, showing the broadband light source, aperture, beam-shaping lenses, sample vial, and spectrometer aligned along the optical axis. (b) Photograph of the complete optical setup on a breadboard, including a zoomed-in view of the sample holder securing the vial at the beam center for stable and repeatable transmission measurements.

*2.2. Measurements*

Figure 2 presents the complete colorimetric sample set and the corresponding raw transmission spectra acquired across a wide range of dye concentrations. Figure 2(a) displays the full set of food dye solutions, prepared by serial dilution from a 1000-unit stock solution. Each concentration was generated by precise volumetric mixing with deionized water and scaled to a final volume of exactly 1 milliliter to maintain uniform optical path length across all vials. This consistency ensured that transmission differences could be attributed solely to dye concentration, not variations in sample geometry.

All spectral measurements were conducted in a dark environment to eliminate ambient light contamination. As mentioned above, to improve robustness while introducing realistic experimental variability, each sample vial was manually repositioned three times within the sample holder. The transmission spectrum was collected after each placement, and the resulting spectra were averaged to obtain a representative curve for each concentration level. A light smoothing filter was applied post-acquisition to reduce high-frequency spectrometer noise without distorting the spectral envelope or relative intensity profile.

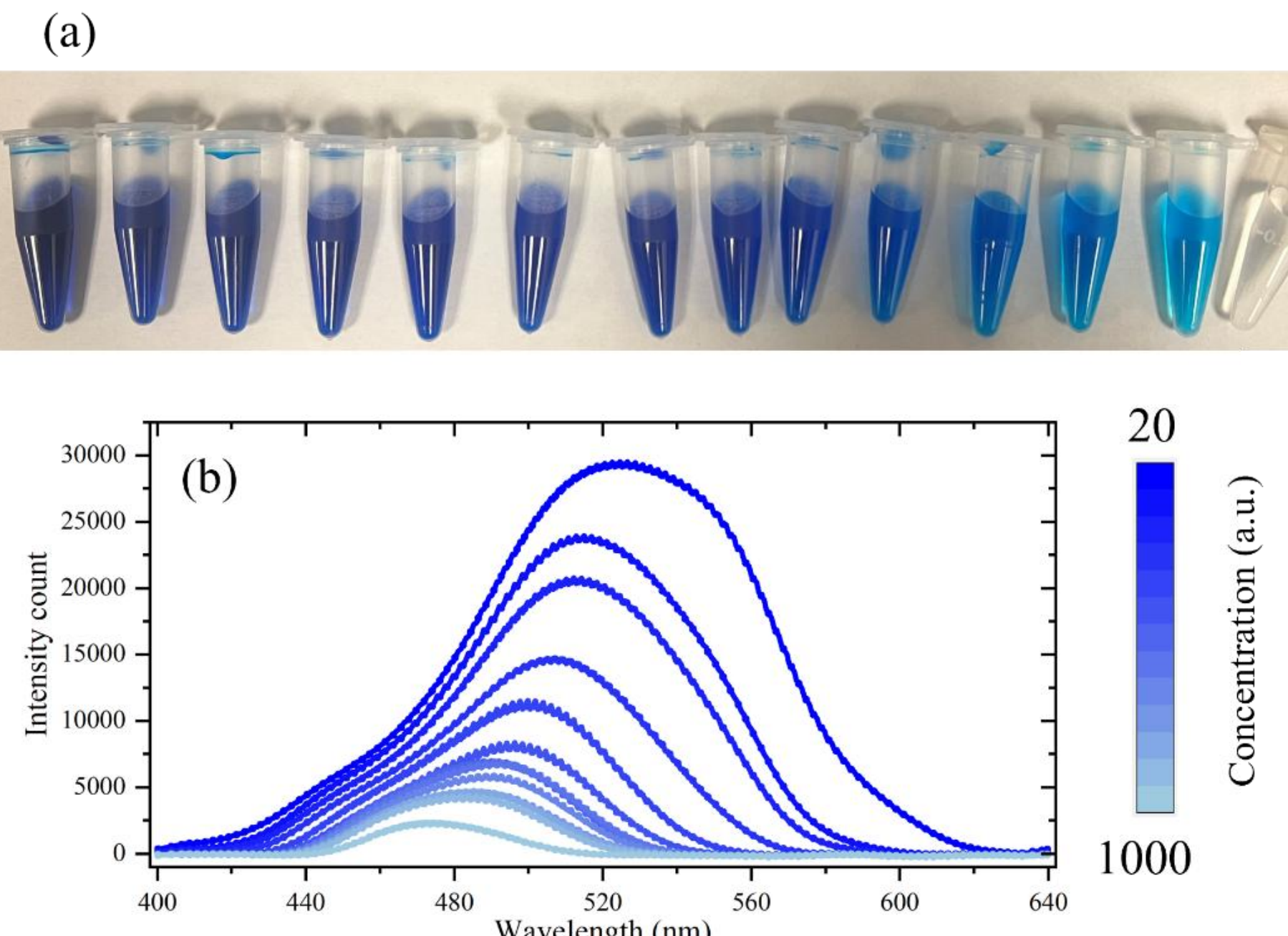

**Figure 2:** (a) Food dye samples ranging from 20 to 1000 concentration units (right to left), prepared by serial dilution from a 1000-unit stock and scaled to 1 mL to ensure consistent optical path length. (b) Raw transmission spectra collected for each concentration, showing decreasing intensity with increasing dye content. Spectra were averaged over three repeated placements per sample and lightly smoothed to reduce high-frequency noise while preserving spectral shape.

Figure 2(b) shows the resulting transmission spectra for concentrations ranging from 20 to 1000 units. As expected, lower concentrations result in higher overall transmission across the visible spectrum due to reduced dye absorption. The spectral shape remains consistent, with a dominant transmission peak near 480-530 nm range and systematic amplitude changes as dye concentration increases. Notably, the visual modulation appears most pronounced in the 500–520 nm range, a region commonly targeted in traditional colorimetric assays. While conventional methods often rely on a single visually selected wavelength, later sections will demonstrate how performance varies dramatically across the spectrum and how greedy feature selection can identify combinations of wavelengths that dramatically improve prediction accuracy of the unknown concentration.

## 3. Results

### *3.1. Single-wavelength Modeling*

While full-spectrum data offers a wealth of information for concentration prediction, most traditional colorimetric systems still operate using just a single measurement wavelength. This is often done out of simplicity, legacy practice, or the assumption that the most visibly modulated part of the spectrum must also be the most predictive. To assess the validity and limitations of this assumption, we carried out a detailed investigation of one-dimensional linear models across a range of individual wavelengths. These models attempt to map transmission intensity at a single fixed wavelength to analyte concentration using one-dimensional linear fitting.

Figure 3 shows a comprehensive panel of such wavelength vs concentration linear fits, spanning from 425 nm to 625 nm. Each subplot displays a different wavelength where transmission values measured across all concentrations, referring back to Fig. 2(b), were used to train a 1-dimensional linear regression model. At first glance, some wavelengths do indeed produce reasonable linear trends. The fits near 450 to 475 nm show relatively tight clustering of points around the fitted line and low RMSE values. However, other wavelengths, even just 25 or 50 nm away, quickly degrade in performance. The wavelength at 550 nm, for instance, yields a significantly higher RMSE than 475 nm, despite still falling within the region of strong visible modulation in the raw spectra. The situation worsens at the spectral tails. Fits at 425 nm and 625 nm show significant nonlinearity in the data, resulting in a weaker linear fit and large prediction errors. Furthermore, the occurrence of negative transmission values at certain wavelengths, which are physically non-meaningful, suggests the presence of measurement noise, baseline drift, or background subtraction errors, further highlighting the reduced reliability and robustness of single-wavelength approaches in these spectral regions.

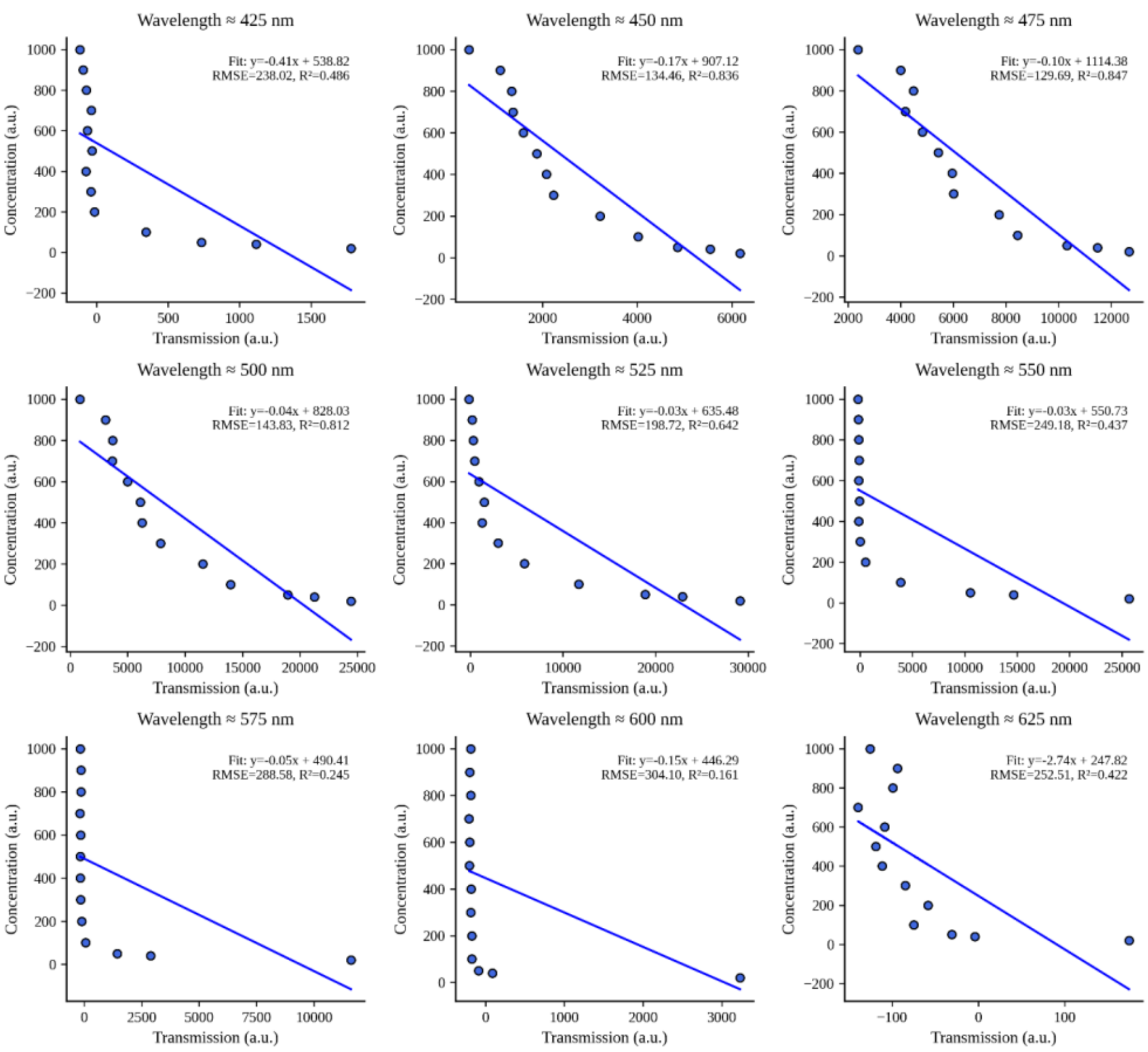

**Figure 3:** Single-wavelength simple linear regression fits at nine selected wavelengths from 425 to 625 nm, showing varying prediction quality and highlighting the limitations of manual wavelength selection.

The key takeaway from Fig. 3 is that even though the underlying spectral shape varies smoothly, as shown in Fig. 2(b), the linear performance across wavelengths does not. The relationship between transmission and concentration is highly sensitive to wavelength choice, and small shifts in the selected value can make the difference between a usable model and a completely unreliable one. This highlights a major limitation of one-dimensional modeling. It depends entirely on a well-chosen wavelength. Unless this wavelength is picked through an objective search or cross-validation, there is no guarantee it will yield meaningful predictions.

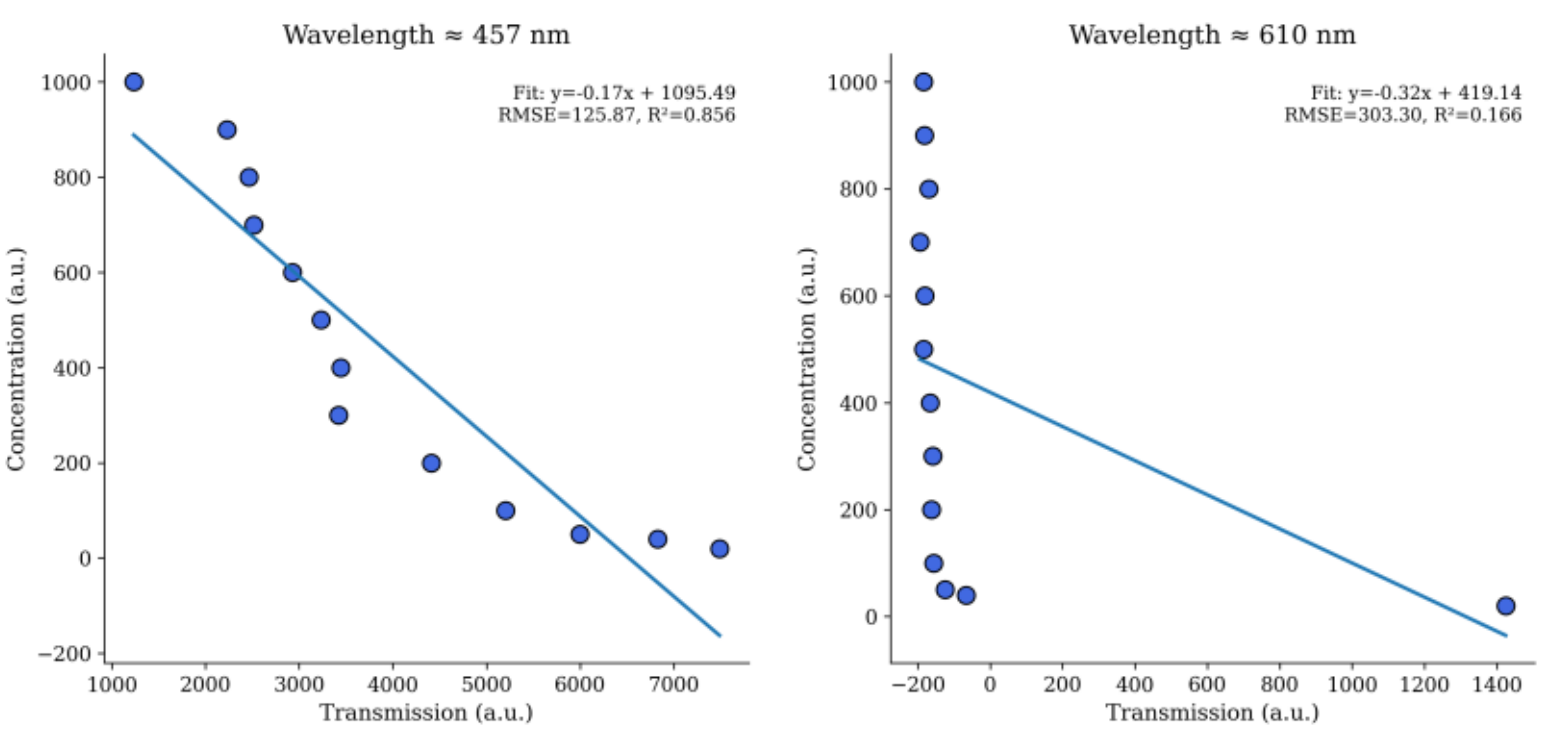

**Figure 4:** (a) Best-performing single-wavelength linear fit at 457 nm with RMSE of 125.87 and $R^2$ of 0.8563. (b) Worst-performing fit at 610 nm with RMSE of 303.30 and $R^2$ of 0.1658, highlighting the non-intuitive nature of optimal wavelength selection.

To probe this issue more closely, Fig. 4 isolates the best and worst performing wavelengths identified from the wavelength sweep. The left panel, Fig. 4(a), shows the simple linear fit at 457 nm, the wavelength with the lowest RMSE. Surprisingly, 457 nm is not visually the most dynamic point on the spectrum. One might have expected that a peak or steep slope region, such as 520 nm or 480 nm, would dominate. Yet statistical evaluation reveals that 457 nm offers the strongest predictive power, even though it may appear relatively unremarkable in the raw transmission profile.

In contrast, Fig. 4(b) shows the linear fit at 610 nm, one of the worst performing wavelengths. The data is widely scattered, includes noise and unrealistic negative values, and the $R^2$ drops to just 0.17.

It is important to note that all results in Figs. 3 and 4 were obtained without cross-validation. That is, the same dataset was used for both training and testing, which results in an idealized best-case scenario fitting performance. This form of evaluation is acceptable for probing the basic relationships between variables but is insufficient to estimate real-world measurement modeling performance. In practice, unknown samples with previously unseen concentrations must be predicted without prior exposure. This scenario must be simulated using proper statistical methods such as k-fold cross-validation, where the model is repeatedly trained on subsets of the data and evaluated on held-out samples.

For our machine learning models, we adopt ten-fold cross-validation to simulate this realistic scenario. The dataset is split into ten equal parts. In each iteration, nine parts are used to train the model, and one part is reserved for testing. This process is repeated ten times so that every data point is tested exactly once. The resulting error metrics are averaged to produce a more generalizable estimate of model performance. Cross-validation not only prevents overfitting but also reveals how well a model can handle variability between different samples, including small deviations in optical alignment or measurement noise. In the context of medical diagnostics, this directly mimics predicting the analyte concentration of a new patient sample with unknown concentration using a trained regression model.

When cross-validation is applied to the single-wavelength models described in Figs. 3 and 4, the performance degrades sharply. For the previously best-performing 457 nm wavelength, MSE jumps from around 15,000 to over 22,000. Even more dramatically, the $R^2$ value, which was originally positive and relatively high, becomes negative under cross-validation, indicating that the model performs worse than simply using the mean as a predictor. For this reason, the regression fit for the cross-validated model is not even plotted, as it offers no meaningful predictive value. This observation alone is enough to illustrate the fragility of one-dimensional fitting in real applications.

This degradation in cross-validated performance of single wavelength fitting stands in stark contrast to the behavior of our multi-feature machine learning models. When trained and tested on the same dataset without cross-validation, the multi-wavelength models achieve near-zero mean squared error and visually perfect fits using only a small number of intelligently selected wavelengths. Although modeling results without any cross-validation are not the focus of this work, they are worth mentioning to emphasize the predictive power that emerges when only multiple wavelengths are leveraged rather than one. In contrast to the one-dimensional models that rapidly lose accuracy under validation, the multi-wavelength machine learning models maintain strong generalization even when subjected to ten-fold cross-validation, which will be presented in detail later in this article. This comparison highlights the fundamental gap between single-wavelength overfitting and robust multi-wavelength modeling.

To further illustrate this point, Fig. 5 presents the RMSE and $R^2$ values for single-wavelength linear regression models across the full spectral range from 400 nm to 640 nm. Each point in these plots corresponds to a regression model trained at one specific wavelength using transmission intensity as the sole input variable.

The top panel of Fig. 5(a) shows the original intensity spectra, repeated here for reference to aid visual comparison with regression performance. These spectra show a clear decay in peak amplitude with increasing concentration, yet they do not inherently reveal which wavelengths yield the best predictive behavior. The middle panel, Fig. 5(b), presents the root mean squared error (RMSE) as a function of wavelength under two evaluation modes: without cross-validation (blue) and with ten-fold cross-validation (green). In the absence of cross-validation, RMSE drops sharply in the 450–480 nm region, forming a clear local minimum where prediction error is lowest. However, this region does not perfectly align with the visually dominant features in the intensity spectra, again reinforcing that statistical relevance does not always coincide with visual salience.

Once ten-fold cross-validation is applied, the green curve in Fig. 5(b) reveals the full impact of generalization constraints. The RMSE curve becomes noisier and shifts upward across nearly the entire spectral range. Even in regions that previously showed strong performance, such as around 457 nm, RMSE increases substantially under validation. This demonstrates that one-dimensional models are highly susceptible to overfitting when tested on the same data they were trained on, and their performance rapidly deteriorates when asked to generalize.

Finally, Fig. 5(c) plots the coefficient of determination ($R^2$) across all wavelengths, again under the no cross-validation scenario. The trend mirrors that of the RMSE plot, peaking around 0.86 near 457 nm and steadily declining away from that region. Past 600 nm, the $R^2$ value drops below 0.2, reflecting minimal predictive value. No $R^2$ curve is shown for the cross-validation case because many of the corresponding values are negative, indicating worse-than-mean prediction and rendering the metric effectively meaningless.

Another key insight from Fig. 5 is that although the raw spectral signal appears smooth and continuous, the actual predictive information is unevenly distributed. There are narrow regions where concentration can be estimated with moderate confidence using a single wavelength, but most of the spectrum offers little usable information in isolation. This strongly supports the case for multivariate modeling. Without a principled method for feature selection or validation, choosing a wavelength manually is nearly indistinguishable from guessing.

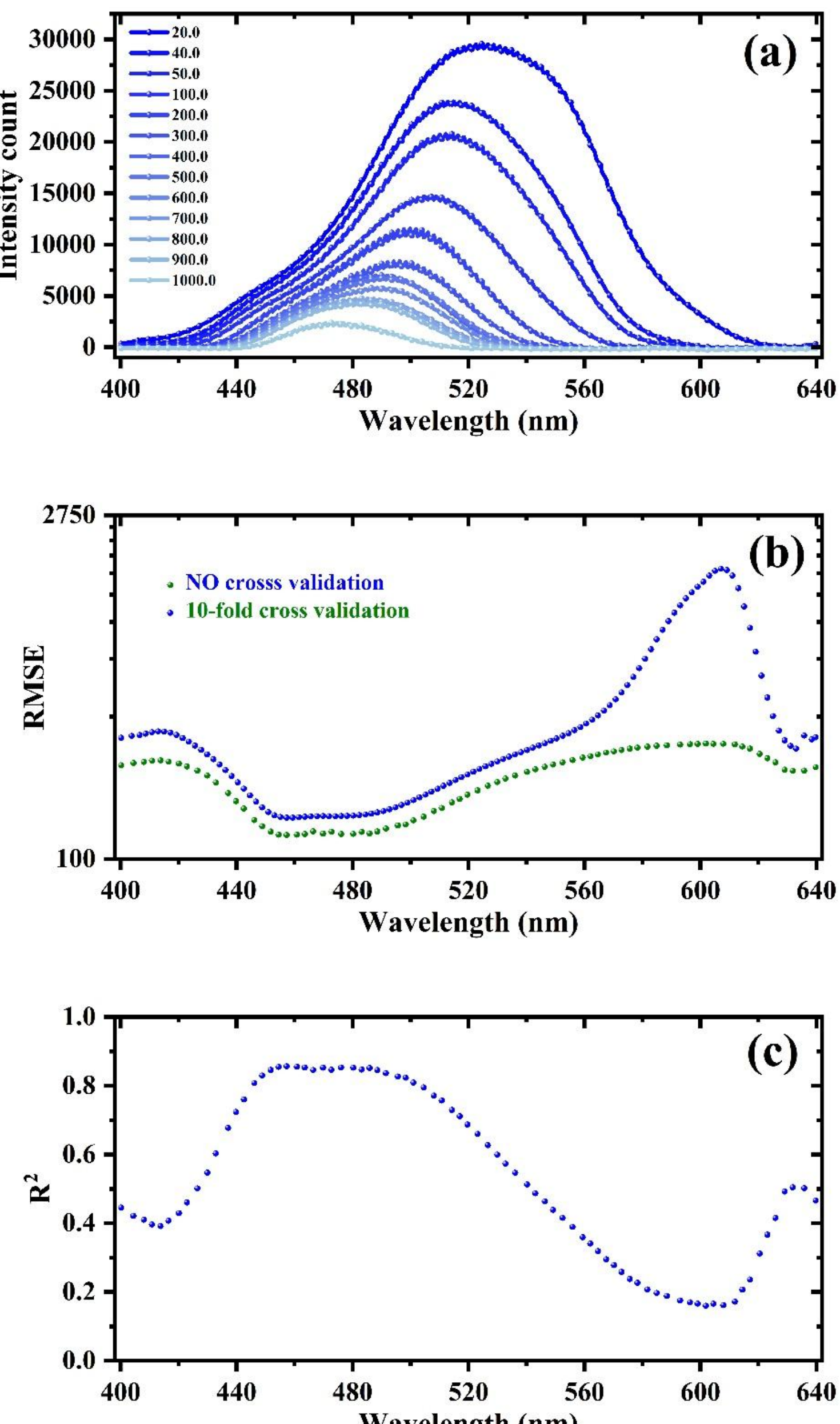

**Figure 5:** (a) Raw transmission spectra across 400–640 nm for all dye concentrations. (b) RMSE trends for single-wavelength linear regression models with and without ten-fold cross-validation. (c) Corresponding $R^2$ values without cross-validation. These plots illustrate the performance variability and generalization gap in single-feature models across the spectrum.

Together, Figs. 3, 4, and 5 paint a consistent picture: single-wavelength colorimetric analysis is highly limited, particularly when prediction must generalize beyond a training set. Even seemingly strong fits collapse under ten-fold cross-validation. This also implies that in most cases measurement inaccuracy is not a result of the sensor hardware limitations, further signifying the importance of using the maximum modeling capacity based on the measured spectra. As previously mentioned, our machine learning models trained without cross-validation achieve zero or near-zero MSE and perfect visual fits using just a few selected wavelengths. These results, although idealized, illustrate the upper bound of what is possible when the spectral data is fully utilized. The rest of this work focuses on the more realistic case of cross-validated multi-feature modeling, which will demonstrate how spectral information, when carefully selected and modeled, can yield robust and generalizable sensing performance and can result in thousands of fold improvement in MSE compared to single-wavelength modeling.

### *3.2. Multi-wavelength modeling*

Table 1 and Fig. 6 together present a comprehensive summary of the greedy forward feature selection process applied to normalized transmission data. This combined analysis marks the transition from single-wavelength fitting to a more intelligent multi-feature modeling strategy, where machine learning is used not only for regression but for optimal information extraction from the spectrum. Greedy forward feature selection iteratively builds a feature subset by adding, at each step, the wavelength that provides the largest improvement in a chosen performance metric (e.g., reduction in MSE), without reconsidering previously selected features [32].

**Table 1.** Stepwise feature selection results showing wavelength identifiers, MSE and RMSE values from 10-fold cross-validation, and fold improvements relative to the 1-feature baseline.

| Total features | feature added in nm (wavelength) | MSE | RMSE | MSE improvement (folds) | RMSE improvement (folds) |
|---|---|---|---|---|---|
| 1 | 457.275 | 22157.58 | 148.8542 | 1 | 1 |
| 2 | 427.16 | 6753.58 | 82.18017 | 3.280864 | 1.811316 |
| 3 | 631.853 | 6861.9 | 82.83659 | 3.229074 | 1.796962 |
| 4 | 423.564 | 4697.08 | 68.53525 | 4.717309 | 2.171937 |
| 5 | 478.19 | 3205.72 | 56.61908 | 6.911889 | 2.629047 |
| 6 | 479.65 | 2388.56 | 48.8729 | 9.276543 | 3.045742 |
| 7 | 545.117 | 1975.39 | 44.44536 | 11.21681 | 3.349151 |
| 8 | 629.068 | 3017.36 | 54.9305 | 7.343366 | 2.709865 |
| 9 | 452.447 | 2759.62 | 52.53209 | 8.029214 | 2.833587 |
| 10 | 547.774 | 8.81 | 2.968164 | 2515.049 | 50.15026 |
| 11 | 631.058 | 18.37 | 4.286024 | 1206.183 | 34.73014 |
| **12** | **488.603** | **3.87** | **1.967232** | **5725.473** | **75.66685** |
| 13 | 455.176 | 32.14 | 5.669215 | 689.4082 | 26.25658 |
| 14 | 633.841 | 70.15 | 8.37556 | 315.86 | 17.77245 |
| 15 | 633.046 | 29.79 | 5.458022 | 743.7925 | 27.27256 |
| 16 | 635.827 | 22.73 | 4.767599 | 974.8165 | 31.22205 |
| 17 | 421.407 | 115.99 | 10.76987 | 191.0301 | 13.82136 |
| 18 | 419.329 | 166.32 | 12.89651 | 133.2226 | 11.54221 |
| 19 | 417.21 | 234.04 | 15.29837 | 94.67433 | 9.730073 |
| 20 | 415.089 | 295.73 | 17.1968 | 74.92503 | 8.655925 |

Table 1 lists the ordered sequence of wavelengths added during the feature selection process, along with the corresponding mean squared error (MSE), root mean squared error (RMSE), and the relative improvement in both metrics compared to the single-wavelength baseline. The first row establishes the baseline case, where only the best-performing single wavelength (457.275 nm) is used, yielding an RMSE of 148.85. As additional wavelengths are incorporated, both MSE and RMSE drop sharply, especially within the first 3 to 5 additions. Notably, by the time twelve features have been selected, the MSE drops to 3.87 and RMSE to just 1.97. This represents a staggering 5725-fold improvement in MSE and over 75-fold improvement in RMSE just by using 12 wavelengths rather than just one wavelength (the best performing wavelength).

Figure 6 offers a visual narrative of this optimization process across six panels. In Fig. 6(a) and Fig. 6(b), the absolute MSE and RMSE values are plotted against the number of selected features. A steep decline is visible in both curves within the first 5 to 7 features, indicating that a large proportion of predictive power is concentrated in a small number of well-chosen wavelengths. MSE drops sharply at the 10-wavelength scenario, and it drops to its lowest at the 12-wavelength fitting.

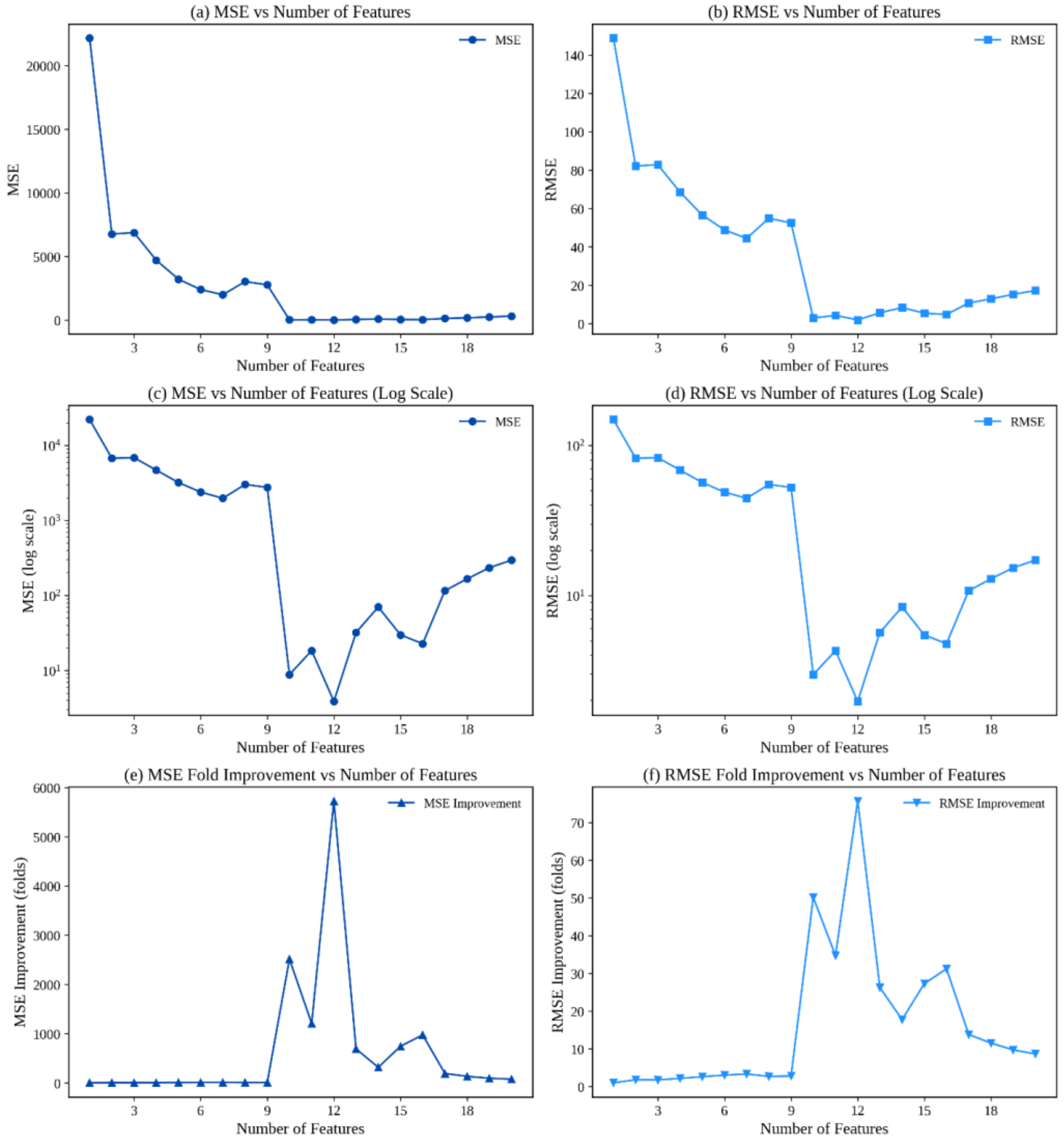

**Figure 6:** Prediction error and enhancement trends as a function of the number of selected features. (a, b) MSE and RMSE values decrease rapidly with added features using normalized transmission and 10-fold cross-validation. (c, d) Corresponding log-scale plots reveal non-linear behavior in prediction error reduction, especially for feature sets between 10 and 15 variables. (e, f) Fold improvement in MSE and RMSE highlights optimal feature ranges where modeling performance gains are maximized.

Panels Fig. 6(c) and Fig. 6(d) replot the MSE and RMSE curves on a logarithmic scale. These log-scale versions emphasize the dramatic nature of the improvement achieved around the tenth feature, where MSE plummets from thousands to single digits. This is especially critical when comparing to the baseline MSE of over 22,000 from single-wavelength fitting.

Finally, Fig. 6(e) and Fig. 6(f) display the fold improvement in MSE and RMSE, respectively. These panels isolate the gain brought by each additional feature, compared to the single-wavelength fitting. The peak in Fig. 6(e) corresponds exactly to the 12th feature, echoing the highlight from Table 1. The curve in Fig. 6(f) tells the same story for RMSE, albeit with a smaller dynamic range. Together, these plots demonstrate that feature number 12 was not merely another increment but a pivotal addition that elevated the model to an entirely different tier of performance.

These findings reinforce that performance gains are not uniformly distributed across the spectral domain. Instead, a small subset of wavelengths carries an outsized share of predictive information, and these can be discovered only through data-driven modeling. Arbitrary or visually chosen wavelengths simply cannot match the predictive power achieved here.

Taken together, Table 1 and Fig. 6 establish both the theoretical and practical value of intelligent feature selection in colorimetric sensing. They show that machine learning models do not need to rely on full-spectrum fitting or brute-force analysis. Instead, they can achieve near-perfect performance through judicious selection of just a dozen highly informative wavelengths. This makes the approach scalable and computationally efficient, opening the door to real-time or embedded implementations without sacrificing accuracy. As such, the results presented here serve as a cornerstone for modernizing traditional colorimetric methods through interpretable and compact machine learning pipelines, especially simple multi-wavelength linear regression models.

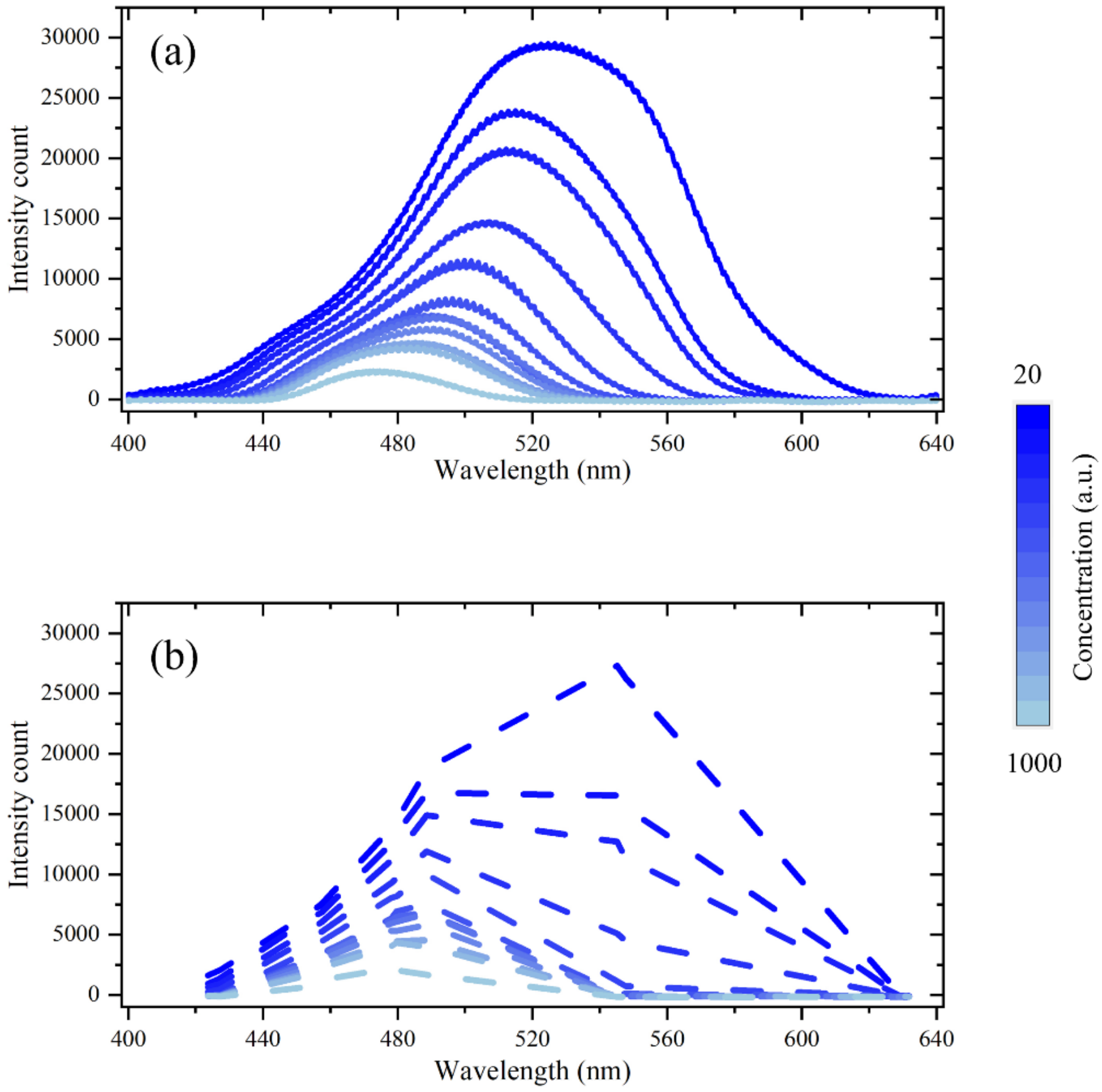

**Figure 7:** (a) Raw transmission spectra of all dye concentrations. (b) Reduced spectra showing the 12 selected features across concentrations. (c) Full-spectrum Pearson correlation heatmap showing broad redundancy. (d) Correlation heatmap of selected wavelengths, confirming low mutual correlation among chosen features.

Figure 7 provides a critical conceptual bridge between raw spectral behavior and data-driven feature selection, offering a more nuanced understanding of why the 12 selected wavelengths performed so well in the previous models. While the prior table and performance plots quantified the benefit of feature addition step by step, this figure turns our attention toward why those particular wavelengths are effective and adds an intuitive understanding.

Figure 7(a) shows the raw intensity spectra for all concentrations across the full wavelength range, just as seen earlier in Fig. 2(b). Figure 7(b), however, transforms those same curves into a representation based on the 12 selected features, in effect, a visual approximation of what the model sees when relying solely on those chosen wavelengths. The dashed lines trace the transmission values at those 12 specific points across all dye concentrations, simulating a sparse sensing regime with high interpretability.

What immediately stands out in Fig. 7(b) is that these 12 wavelengths preserve the overall shape and concentration-dependent structure of the full spectra remarkably well. Even though the resolution is drastically reduced, the essential contrast between high and low concentration samples is maintained with clarity. This preservation of signal shape and ranking, despite aggressive dimensionality reduction, is the conceptual key to why the selected wavelengths work so well. The model is not randomly guessing, it is identifying anchor points in the spectrum that jointly approximate the full signal's behavior. In short, Fig. 7 helps explain the why behind the what. The selected wavelengths work well not only because they reduce MSE numerically, but because they are physically distributed across the spectrum in a way that captures concentration-induced modulation without falling into the trap of redundancy.

## 4. Conclusions

This work establishes an experimental and analytical framework for optimized multi-wavelength colorimetric sensing. Through a deliberately simple yet rigorously validated optical setup, we demonstrate that full-spectrum transmission data, when properly modeled, can achieve levels of accuracy that are often presumed to require sophisticated instrumentation. Although one-dimensional regression initially appeared promising at specific wavelengths such as 457 nm, this performance proved fragile. Without cross-validation, models showed acceptable fits ($R^2 \approx 0.86$), but under ten-fold cross-validation, used here as a proxy for real-world variability, performance deteriorated sharply. In many cases, $R^2$ became negative and RMSE exceeded 30,000, revealing a fundamental limitation of the single-wavelength approach.

To address this, we adopted a multi-wavelength framework based on greedy forward feature selection and ridge regression. Using interpretable linear models on normalized transmission spectra, we found that just twelve carefully selected wavelengths reduced the mean squared error from over 22,000 to 3.87 under 10-fold cross-validation, corresponding to a 5,700-fold improvement without any hardware modification. The selected wavelengths were distributed across the spectrum, capturing complementary, non-redundant information and forming a compact, physically meaningful basis for concentration estimation. These results challenge the assumption that visually dominant wavelengths are the most informative, instead showing that predictive power lies in distributed spectral features. More broadly, this work reframes colorimetric sensing as a high-dimensional data problem and supports the hypothesis that linear regression is well suited for intensity-based sensing, where the relationship with analyte concentration remains quasi-linear for a given wavelength.

**Author Contributions:** For research articles with several authors, a short paragraph specifying their individual contributions must be provided. The following statements should be used "Conceptualization, M.A., A.T. and X.F.; methodology, M.A and C.R.; software, M.A. and C.R.; validation, M.A.; formal analysis, M.A.; investigation, M.A.; resources, X.F.; data curation, M.A. and C.R.; writing—original draft preparation, M.A.; writing—review and editing, M.A., C.R., A.T. and X.F;

visualization, M.A., C.R. and X.F.; supervision, X.F.; project administration, X.F.; funding acquisition, X.F. All authors have read and agreed to the published version of the manuscript.

**Funding:** This research was funded by the National Science Foundation through grant number 2225568.

**Data Availability Statement:** The raw data supporting the conclusions of this article will be made available by the authors on request.

**Conflicts of Interest:** The authors declare no conflicts of interest.